\documentclass[aps,prc,preprint,amsmath,amssymb,showpacs,preprintnumbers,superscriptaddress]{revtex4-1}
\usepackage{CJK}
\usepackage{graphicx}
\usepackage{dcolumn}
\usepackage{bm}
\usepackage{tabularx}

\begin{document}

\title{Multi-chiral facets in symmetry restored states: Five chiral doublets candidates in even-even nucleus $^{136}$Nd}
\author{Y. K. Wang}
\affiliation{State Key Laboratory of Nuclear Physics and Technology, School of Physics, Peking University, Beijing 100871, China}
\author{F. Q. Chen}
\affiliation{School of Natural and Applied Sciences, Northwestern Polytechnical University, Xian 710129, China}
\author{P. W. Zhao}
\affiliation{State Key Laboratory of Nuclear Physics and Technology, School of Physics, Peking University, Beijing 100871, China}
\author{S. Q. Zhang}
\affiliation{State Key Laboratory of Nuclear Physics and Technology, School of Physics, Peking University, Beijing 100871, China}
\author{J. Meng}\email{mengj@pku.edu.cn}
\affiliation{State Key Laboratory of Nuclear Physics and Technology, School of Physics, Peking University, Beijing 100871, China}
\affiliation{School of Physics and Nuclear Energy Engineering, Beihang University, Beijing 100191, China}
\affiliation{Yukawa Institute for Theoretical Physics, Kyoto University, Kyoto 606-8502, Japan}

\date{\today}
\begin{abstract}
  A triaxial projected shell model including configurations with more than four quasiparticles in the configuration space is developed, and applied to investigate the recently reported five chiral doublets candidates in a single even-even nucleus $^{136}$Nd.
  The energy spectra and transition probability ratios $B(M1)/B(E2)$ are reproduced satisfactorily.
  The configuration mixing along the rotational bands is studied by analyzing the intrinsic composition of the eigenfunctions.
  The chiral geometry of these nearly degenerate bands is examined by the \textit{K plot} and the \textit{azimuthal plot}, and the evolution from the chiral vibration to the static chirality with spin is clearly demonstrated for four pairs of partner bands.
  From the features in the \textit{azimuthal plot}, it is difficult to interpret the other candidate as chiral partners.
\end{abstract}
\maketitle
\date{\today}

Chirality is one of the hot topics in biology, chemistry, and physics.
The chirality in atomic nuclei was first predicted theoretically by Frauendorf and Meng in 1997~\cite{Frauendorf1997Nucl.Phys.A617131--147}.
The predicted topology, the mutually perpendicular angular momenta of the valence protons, valence neutrons, and the core, leads to the so-called spontaneous chiral symmetry breaking in the intrinsic frame.
The restoration of the chiral symmetry in the laboratory frame can be manifested by the observation of a pair of nearly degenerate $\Delta I = 1$ bands with the same parity.

Within the framework of the adiabatic and configuration fixed constrained triaxial covariant density
functional theory (CDFT)~\cite{meng2016relativistic}, a new phenomenon, the existence of multiple chiral doublets~(abbreviated as M$\chi$D), i.e., more than one pair of chiral doublet bands in one single nucleus has been predicted~\cite{Meng2006Phys.Rev.C73037303,Peng2008Phys.Rev.C77024309,Yao2009Phys.Rev.C79067302,Li2011Phys.Rev.C83037301}.

Many experimental and theoretical efforts have been devoted to study chiral doublets in various mass regions.
So far, around 60 pairs of chiral doublets candidates have been established in $A \sim 80$~\cite{Wang2011PhysicsLettersB70340-45,Liu2016Phys.Rev.Lett.116112501}, $100$~\cite{Vaman2004Phys.Rev.Lett.92032501,Joshi2007Phys.Rev.Lett.98102501,Tonev2014Phys.Rev.Lett.112052501,
Lieder2014Phys.Rev.Lett.112202502,Rather2014Phys.Rev.Lett.112202503}, $130$~\cite{Starosta2001Phys.Rev.Lett.86971--974,Zhu2003Phys.Rev.Lett.91132501,Tonev2006Phys.Rev.Lett.96052501,
Petrache2006Phys.Rev.Lett.96112502,Grodner2006Phys.Rev.Lett.97172501,Mukhopadhyay2007Phys.Rev.Lett.99172501,
Ayangeakaa2013Phys.Rev.Lett.110172504}, and $190$~\cite{Balabanski2004Phys.Rev.C70044305,Lawrie2008Phys.Rev.C78021305} mass regions.
The observation of chiral bands in different mass regions indicates that nuclear chirality could be universal.
The experimental evidences for the M$\chi$D were reported in $^{133}$Ce~\cite{AyangeakaaPhys.Rev.Lett.110172504}, $^{103}$Rh~\cite{Kuti2014Phys.Rev.Lett.113032501}, and $^{78}$Br~\cite{Liu2016Phys.Rev.Lett.116112501} which demonstrate the multiple facets of nuclear chirality.
For the detailed experimental status, see the data compilation~\cite{Xiong2019AtomicDataandNuclearDataTables125193-225}.

Theoretically, the chiral doublet bands have been investigated by various approaches, for example, the particle rotor model (PRM)~\cite{Frauendorf1997Nucl.Phys.A617131--147,Peng2003Phys.Rev.C68044324,Koike2004Phys.Rev.Lett.93172502,
Zhang2007Phys.Rev.C75044307,Qi2009PhysicsLettersB675175-180}, the tilted axis cranking model (TAC)~\cite{Frauendorf1997Nucl.Phys.A617131--147,Dimitrov2000Phys.Rev.Lett.845732--5735,Olbratowski2004Phys.Rev.Lett.93052501,Zhao2017PhysicsLettersB7731-5}, the TAC approach with the random phase approximation~\cite{Almehed2011Phys.Rev.C83054308} and the collective Hamiltonian~\cite{Chen2013Phys.Rev.C87024314,Chen2016Phys.Rev.C94044301}, the interacting boson-fermion-fermion model~\cite{Tonev2006Phys.Rev.Lett.96052501,Tonev2007Phys.Rev.C76044313,Brant2008Phys.Rev.C78034301}, the generalized coherent state model~\cite{Raduta2016JournalofPhysicsG:NuclearandParticlePhysics43095107} and the projected shell model (PSM)~\cite{Bhat2012PhysicsLettersB707250-254, Chen2017Phys.Rev.C96051303,Chen2018PhysicsLettersB785211-216}.

The PSM carries out the shell model configuration mixing based on Nilsson mean field with the angular momentum projection technique~\cite{HARA1995InternationalJournalofModernPhysicsE04637-785}.
It has been successfully used to investigate the backbending phenomena~\cite{Hara1991ZeitschriftfurPhysikAHadronsandNuclei33915--21}, superdeformed rotational bands~\cite{Sun1999Phys.Rev.Lett.83686--689}, signature inversion~\cite{Gao2006PhysicsLettersB634195-199}, and $\gamma$ bands~\cite{Sheikh2008Phys.Rev.C77034313,Sheikh2010PhysicsLettersB688305-308}, etc.
Recently, this idea has also been implemented based on the self-consistent relativistic~\cite{Zhao2016Phys.Rev.C94041301} and nonrelativistic~\cite{Konieczka2018Phys.Rev.C97034310} density functional theories.

The PSM was first adopted to investigate the chiral rotation in Cs isotopes by Bhat \textit{et al.}~\cite{Bhat2012PhysicsLettersB707250-254}, where the energy spectra and electromagnetic transitions of chiral bands are well reproduced.
However, the underlying chiral geometry was failed to be illustrated due to the fact that the angular momentum geometry is defined in the intrinsic frame.
By introducing the \textit{K plot} and the \textit{azimuthal plot}, Chen \textit{et al.} illustrated the evolution from chiral vibration to static chiral rotation in the framework of PSM~\cite{Chen2017Phys.Rev.C96051303,Chen2018PhysicsLettersB785211-216}.
As a result, the PSM has become a powerful tool to study the chiral rotation.

Recently, five pairs of nearly degenerate bands were observed in the even-even nucleus $^{136}$Nd~\cite{Petrache2018Phys.Rev.C97041304}, in which the energy spectra, total angular momenta, and $B(M1)/B(E2)$ ratios are described with the microscopic and self-consistent tilted axis cranking CDFT~\cite{Peng2008Phys.Rev.C78024313, Zhao2011Phys.Lett.B699181--186,Zhao2011Phys.Rev.Lett.107122501,Meng2013Front.Phys.855--79,Zhao2015Phys.Rev.Lett.115022501, meng2016relativistic, Zhao2017PhysicsLettersB7731-5,Zhao2018InternationalJournalofModernPhysicsE271830007}. Of course, the descriptions of the energy splitting and the quantum tunneling between the partner bands are beyond the mean-field approach.
In Ref.~\cite{Chen2018PhysicsLettersB782744-749}, the PRM has been employed to investigate the chiral doublets candidates in $^{136}$Nd.
The energy splitting, quantum tunneling and the chiral interpretation for the five pairs of doublets are given.

In this paper, a triaxial PSM including configurations with more than four quasiparticles in the configuration space is developed, and applied to investigate the five chiral doublets candidates in the even-even nucleus $^{136}$Nd.
The energy spectra and $B(M1)/B(E2)$ ratios are calculated and compared with the data, and the \textit{K plot} and \textit{azimuthal plot} are provided to understand the chiral geometry.

%----------------------------------------------------
The present PSM is based on the following pairing plus quadrupole Hamiltonian~\cite{Ring2004},
\begin{equation}\label{eq1}
  \hat{H} = \hat{H}_0 - \frac{\chi}{2}\sum_{\mu}\hat{Q}^\dag_\mu\hat{Q}_\mu - G_M\hat{P}^\dag\hat{P} - G_Q\sum_{\mu}\hat{P}^\dag_\mu\hat{P}_\mu,
\end{equation}
in which $\hat{H}_0$ is the spherical single-particle shell model Hamiltonian.
The second term is the quadrupole-quadrupole interaction and the last two terms are the monopole and quadrupole pairing interactions, respectively.
The intrinsic vacuum state $|\Phi_0\rangle$ is provided by the solution of the variational equation,
\begin{equation}\label{eq2}
  \delta\langle\Phi_0|\hat{H}-\lambda_n\hat{N}-\lambda_p\hat{Z}|\Phi_0\rangle = 0,
\end{equation}
where the Lagrangian multipliers $\lambda_n$ and $\lambda_p$ are determined by the neutron number $N$ and proton number $Z$, respectively.

Based on the obtained vacuum $|\Phi_0\rangle$, the intrinsic multi-quasiparticle (qp) states $|\Phi_\kappa\rangle$ can be constructed.
For example, the multi-qp configurations up to 6-qp states for even-even nuclei are,
\begin{equation}\label{eq3}
  \begin{split}
    \{|\Phi_0\rangle, &\hat{\beta}^\dag_{\nu_i}\hat{\beta}^\dag_{\nu_j}|\Phi_0\rangle,  \hat{\beta}^\dag_{\pi_i}\hat{\beta}^\dag_{\pi_j}|\Phi_0\rangle, \hat{\beta}^\dag_{\pi_i}\hat{\beta}^\dag_{\pi_j}
    \hat{\beta}^\dag_{\nu_k}\hat{\beta}^\dag_{\nu_l}|\Phi_0\rangle,\\
    &\hat{\beta}^\dag_{\nu_i}\hat{\beta}^\dag_{\nu_j}\hat{\beta}^\dag_{\nu_k}\hat{\beta}^\dag_{\nu_l}|\Phi_0\rangle,
    \hat{\beta}^\dag_{\pi_i}\hat{\beta}^\dag_{\pi_j}\hat{\beta}^\dag_{\pi_k}\hat{\beta}^\dag_{\pi_l}|\Phi_0\rangle,\\
    &\hat{\beta}^\dag_{\nu_i}\hat{\beta}^\dag_{\nu_j}\hat{\beta}^\dag_{\nu_k}\hat{\beta}^\dag_{\nu_l}\hat{\beta}^\dag_{\nu_m}
    \hat{\beta}^\dag_{\nu_n}|\Phi_0\rangle, \hat{\beta}^\dag_{\pi_i}\hat{\beta}^\dag_{\pi_j}\hat{\beta}^\dag_{\pi_k}
    \hat{\beta}^\dag_{\pi_l}\hat{\beta}^\dag_{\pi_m}\hat{\beta}^\dag_{\pi_n}|\Phi_0\rangle,\\
    &\hat{\beta}^\dag_{\pi_i}\hat{\beta}^\dag_{\pi_j}\hat{\beta}^\dag_{\nu_k}\hat{\beta}^\dag_{\nu_l}\hat{\beta}^\dag_{\nu_m}
    \hat{\beta}^\dag_{\nu_n}|\Phi_0\rangle,\hat{\beta}^\dag_{\nu_i}\hat{\beta}^\dag_{\nu_j}\hat{\beta}^\dag_{\pi_k}
    \hat{\beta}^\dag_{\pi_l}\hat{\beta}^\dag_{\pi_m}\hat{\beta}^\dag_{\pi_n}|\Phi_0\rangle\},
  \end{split}
\end{equation}
where $\hat{\beta}^\dag_\pi, \hat{\beta}^\dag_\nu$ ($\hat{\beta}_\pi, \hat{\beta}_\nu$) represent the qp creation (annihilation) operators of proton and neutron, respectively.

The violated rotational symmetry in the intrinsic multi-qp states $|\Phi_\kappa\rangle$ is restored by projection
\begin{equation}\label{eq4}
  \{\hat{P}^I_{MK}|\Phi_\kappa\rangle\},
\end{equation}
in which $\hat{P}^I_{MK}$ is the three dimensional angular momentum projection operator~\cite{Ring2004},
\begin{equation}\label{eq5}
  \hat{P}^I_{MK} = \frac{2I+1}{8\pi^2}\int d\Omega D^{I\ast}_{MK}(\Omega)\hat{R}(\Omega).
\end{equation}
The pairing plus quadrupole Hamiltonian is then diagonalized in the projected basis thus obtained,
\begin{equation}\label{eq6}
  \sum_{\kappa'K'}\{\langle\Phi_{\kappa}|\hat{H}\hat{P}^I_{KK'}|\Phi_{\kappa'}\rangle-E^{I\sigma}\langle\Phi_{\kappa}|\hat{P}^I_{KK'}|
  \Phi_{\kappa'}\rangle\}f^{I\sigma}_{K'\kappa'} = 0,
\end{equation}
in which $\sigma$ specifies different eigenstates of the same spin $I$.
The norm matrix element $\mathcal{N}_I(K,\kappa; K',\kappa')=\langle\Phi_\kappa|\hat{P}^I_{KK'}|\Phi_{\kappa'}\rangle$ and
the energy kernel $\langle\Phi_{\kappa}|\hat{H}\hat{P}^I_{KK'}|\Phi_{\kappa'}\rangle$ can be calculated with the efficient Pfaffian algorithm~\cite{Hu2014PhysicsLettersB734162-166,Bertsch2012Phys.Rev.Lett.108042505}.

By solving the Hill-Wheeler equation \eqref{eq6}, one can get the eigenvalues $E^{I\sigma}$ and the corresponding eigenfunctions
\begin{equation}\label{eq7}
  |\Psi^\sigma_{IM}\rangle = \sum_{K\kappa} f^{I\sigma}_{K\kappa}\hat{P}^I_{MK}|\Phi_\kappa\rangle,
\end{equation}
and the physical observables, such as transition probabilities $B(M1)$ and $B(E2)$, can then be calculated.

For each eigenfunction $|\Psi_{IM}^\sigma\rangle$, the composition of intrinsic multi-qp state $|\Phi_\kappa\rangle$ can be calculated by its weight
\begin{equation}\label{eq8}
  W_\kappa = \sum_{K}|g^{I\sigma}(K,\kappa)|^2,
\end{equation}
where the orthogonal and normalized collective wavefunctions $g^{I\sigma}(K,\kappa)$~\cite{Chen2017Phys.Rev.C96051303} are
\begin{equation}\label{eq9}
  g^{I\sigma}(K,\kappa) = \sum_{K'\kappa'}\mathcal{N}_I^{1/2}(K,\kappa; K',\kappa')f^{I\sigma}_{K'\kappa'}.
\end{equation}

The collective wavefunctions $g^{I\sigma}(K,\kappa)$ can be used to construct the \textit{K plot}~\cite{Chen2017Phys.Rev.C96051303}, i.e., the probability distribution of the angular momentum components in the intrinsic frame,
\begin{equation}
  p^{I\sigma}(|K|) = \sum_\kappa |g^{I\sigma}(K,\kappa)|^2 + |g^{I\sigma}(-K,\kappa)|^2.
\end{equation}

The \textit{azimuthal plot} depicting the probability distribution profile for the orientation of the angular momentum on the $(\theta,\phi)$~plane in the intrinsic frame, is defined as~\cite{Chen2017Phys.Rev.C96051303}
\begin{equation}
  \mathcal{P}(\theta,\phi) = \sum_\kappa\int_0^{2\pi} d\psi'|G^{II}(\psi',\theta,\pi-\phi,\kappa)|^2,
\end{equation}
where $(\theta,\phi)$ are the tilted angles of the angular momentum with respect to
the intrinsic frame.
The collective wavefunction $G^{II}(\psi',\theta,\pi-\phi,\kappa)$ is as follows~\cite{Chen2017Phys.Rev.C96051303},
\begin{equation}
  G^{II}(\psi',\theta,\pi-\phi,\kappa) = \sqrt{\frac{2I+1}{8\pi^2}}\sum_K g^I(K,\kappa)D^{I\ast}_{IK}(\psi',\theta,\pi-\phi).
\end{equation}

%------------------------------------------------------------------
The five chiral doublets candidates (bands D1-D5 and their partners) observed in $^{136}$Nd~\cite{Petrache2018Phys.Rev.C97041304} are investigated by the PSM.
The strengths of monopole and quadrupole pairing forces in the Hamiltonian \eqref{eq1} are the same as in Ref.~\cite{Sheikh2009NuclearPhysicsA82458-69}.
Three major shells $(N = 3, 4, 5)$ of the Nilsson energy levels are adopted for both protons and neutrons.
The strength of the quadrupole force $\chi$ is determined by the quadrupole deformation parameters $(\beta,\gamma)$ and the corresponding quadrupole moments~\cite{HARA1995InternationalJournalofModernPhysicsE04637-785}.
The quadrupole deformation parameters $(\beta,\gamma)$ adopted in the present calculations are listed in Table~\ref{table1} together with the configurations for each pair of the bands.
The deformation parameters $\beta$ are taken from Ref.~\cite{Petrache2018Phys.Rev.C97041304}, where the self-consistent CDFT calculations were carried out.
The deformation parameters $\gamma$ given by the CDFT calculations vary from $19^\circ$ to $23^\circ$ for the five configurations.
Considering the fact that the potential energy surface of $^{136}$Nd is very soft in the $\gamma$ direction, the deformation parameters
$\gamma$ are fixed to be $30^\circ$ for all configurations.
%-----------------------------------------------------------------------------------------------------
\begin{table*}[!]
  \centering
  \caption{The parities, quadrupole deformation parameters $\beta$ and $\gamma$, qp excitation energies $E_x$, configurations ($\nu$ represents neutron and $\pi$ represents proton), degeneracy of configurations, and the number of integral grids for Euler angles used in the PSM for bands D1-D5 and their partners.}
  \label{table1}
  \begin{tabular}{ccccccc}
    \hline\hline
    Band& Parity& $(\beta,\gamma)$  & $E_x$ & Configuration  &  Degeneracy& Integral \\[-0.3cm]
         &       &                   & (MeV) &               &        & grids     \\
    \hline
    D1   & $+$     & $(0.21,30^\circ)$ & 3.560&$\nu[h_{11/2};9/2][d_{3/2};1/2]\otimes\pi[h_{11/2};3/2][g_{7/2};5/2]$& 8& 7168\\
         &       &                   & 3.570&$\nu[h_{11/2};9/2][d_{3/2};1/2]\otimes\pi[h_{11/2};1/2][g_{7/2};5/2]$& 8&     \\
         &       &                   & 3.578&$\nu[h_{11/2};9/2][d_{3/2};1/2]\otimes\pi[h_{11/2};3/2][g_{7/2};3/2]$& 8&     \\
         &       &                   & 3.588&$\nu[h_{11/2};9/2][d_{3/2};1/2]\otimes\pi[h_{11/2};1/2][g_{7/2};3/2]$& 8&     \\
    \hline
    D2   &$+$      & $(0.22,30^\circ)$ & 6.409&$\nu[h_{11/2};9/2][d_{3/2};1/2]\otimes$                                   &32& 9216\\
         &       &                   &      &$\pi[h_{11/2};1/2][h_{11/2};3/2][h_{11/2};5/2][g_{7/2};5/2]$         &  &     \\
    \hline
    D3, D4&$-$    & $(0.21,30^\circ)$ & 3.688 &$\nu[h_{11/2};9/2][d_{3/2};1/2]\otimes\pi[h_{11/2};1/2][h_{11/2};3/2]$&8& 7168\\
          &      &                   & 4.155 &$\nu[h_{11/2};9/2][d_{3/2};1/2]\otimes\pi[h_{11/2};3/2][h_{11/2};5/2]$&8&     \\
          &      &                   & 4.166 &$\nu[h_{11/2};9/2][d_{5/2};3/2]\otimes\pi[h_{11/2};1/2][h_{11/2};5/2]$&8&     \\
          &      &                   & 4.298 &$\nu[h_{11/2};7/2][d_{3/2};1/2]\otimes\pi[h_{11/2};1/2][h_{11/2};3/2]$&8&     \\
    \hline
    D5    &$+$      & $(0.26,30^\circ)$ & 4.086 &$\nu[h_{11/2};9/2][f_{7/2};1/2]\otimes\pi[h_{11/2};1/2][h_{11/2};3/2]$&8& 9216\\
          &      &                   & 4.155 &$\nu[h_{11/2};9/2][f_{7/2};1/2]\otimes\pi[h_{11/2};3/2][h_{11/2};5/2]$&8&     \\
          &      &                   & 4.335 &$\nu[h_{11/2};9/2][h_{9/2};1/2]\otimes\pi[h_{11/2};1/2][h_{11/2};3/2]$&8&     \\
          &      &                   & 4.364 &$\nu[h_{11/2};9/2][h_{9/2};1/2]\otimes\pi[h_{11/2};3/2][h_{11/2};5/2]$&8&     \\
    \hline\hline
  \end{tabular}
\end{table*}
%------------------------------------------------------------------------------------------

With the quadrupole deformations $(\beta,\gamma)$, the intrinsic vacuum $|\Phi_0\rangle$ in Eq.~\eqref{eq2} and the corresponding intrinsic multi-qp states $|\Phi_\kappa\rangle$ in Eq.~\eqref{eq3} are obtained.
From $|\Phi_0\rangle$ and $|\Phi_\kappa\rangle$, the projected basis $\{\hat{P}^I_{MK}|\Phi_\kappa\rangle\}$ can be constructed.
In order to optimize the computation, the configuration space spanned by the projected basis $\{\hat{P}^I_{MK}|\Phi_\kappa\rangle\}$ is built based on the configurations assigned in Ref.~\cite{Petrache2018Phys.Rev.C97041304}.
Due to the time reversal symmetry of the single quasiparticle state,
a $n$-qp configuration has $2^n/2$ degeneracy as shown in the sixth column of Table~\ref{table1}.
The truncation of the configuration space is confirmed by the convergence of the obtained energy spectra for the $4$-qp bands.

For the negative-parity $4$-qp bands D3 and D4, the assigned configurations in Ref.~\cite{Petrache2018Phys.Rev.C97041304} contain two $h_{11/2}$ quasi-protons and one $h_{11/2}$ quasi-neutron at almost the same deformation.
Therefore, same configuration space shown in Table~\ref{table1} is adopted and the calculated results for bands D3 and D4 together with their partners are obtained simultaneously.

For the positive-parity $4$-qp bands, with the assigned configurations in Ref.~\cite{Petrache2018Phys.Rev.C97041304}, namely, $\nu(h_{11/2})(d_{3/2}) \otimes\pi(h_{11/2})(g_{7/2})$ for band D1, and $\nu(h_{11/2})(h_{9/2}) \otimes\pi(h_{11/2})^2$ for band D5, configuration spaces shown in Table~\ref{table1} are adopted.
For the positive-parity $6$-qp band D2, as shown in Table~\ref{table1}, $32$ degenerate states with configuration assigned in Ref.~\cite{Petrache2018Phys.Rev.C97041304} are used to build the configuration space in present calculations.

The integrals of the norm and energy kernel over the Euler angles $(\Omega = \psi', \theta', \phi')$ are carried out in grids.
It turns out that $16\times14\times32 = 7168$ integral grids can provide convergent energy spectra for bands D1, D3, and D4.
For bands D2 and D5, $16\times18\times32 = 9216$ integral grids are needed.

%-----------------------------------------------------------------------------------
\begin{figure*}[!]
  \centering
  \includegraphics[width=\textwidth]{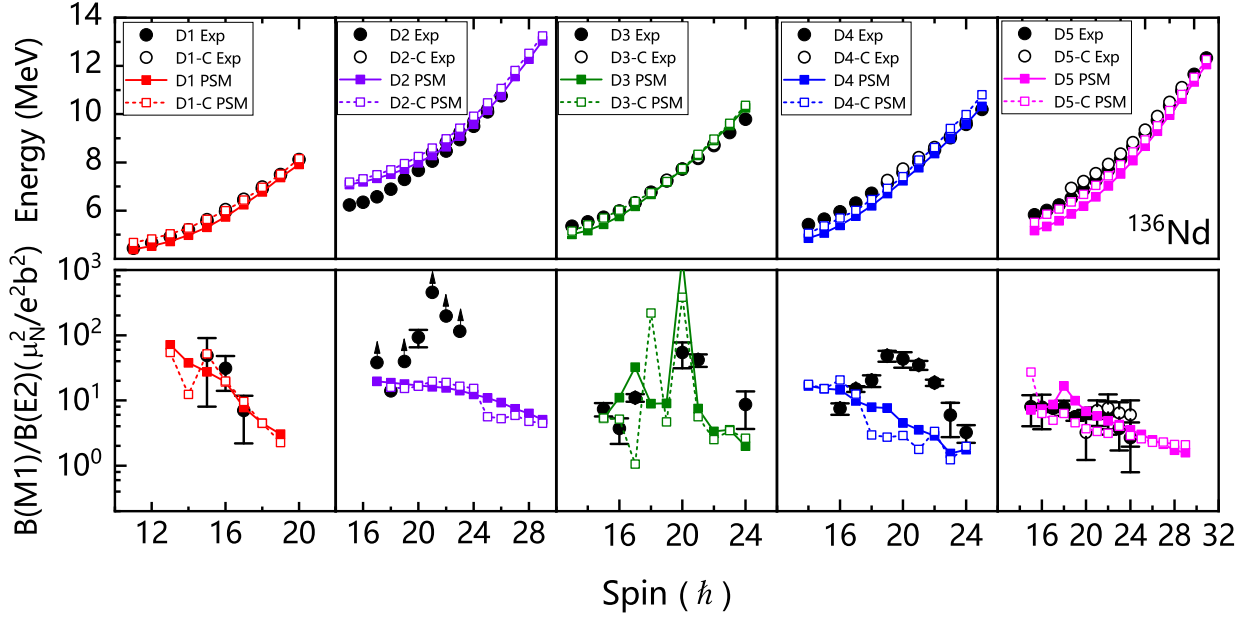}
  \caption{(Color online) Energy spectra and $B(M1)/B(E2)$ ratios calculated by the PSM as functions of angular momenta, in comparison with the experimental data~\cite{Petrache2018Phys.Rev.C97041304}.}
  \label{fig1}
\end{figure*}

In Fig.~\ref{fig1}, the calculated energy spectra and $B(M1)/B(E2)$ ratios with the PSM for bands D1-D5 and their partners are shown in comparison with the available data~\cite{Petrache2018Phys.Rev.C97041304}.
The bandhead of D1 is taken as reference.
The experimental energy spectra and $B(M1)/B(E2)$ ratios for bands D1-D5 are reproduced satisfactorily by the calculations.
For each pair of doublets, the calculated energy spectra are nearly degenerate and the corresponding $B(M1)/B(E2)$ ratios are in general similar to each other.
This is one of the characteristics for the chiral rotation.

For bands D1 and D1-C, in consistent with the experimental data, the energy differences between the doublets are nearly unchanged, and the $B(M1)/B(E2)$ ratios decline with spin.

For bands D2 and D2-C, the energy spectra at lower spins are overestimated, and the bump of $B(M1)/B(E2)$ ratios around $I = 21\hbar$ can not be reproduced by the PSM calculations.
These discrepancies might be due to the neglect of the 4-qp configurations in the present configuration space.
The possible mixing of the 4-qp configurations with 6-qp configurations is also supported by the qp alignments in Ref.~\cite{Petrache2018Phys.Rev.C97041304}.

For bands D3 and D4 together with their partners, the calculated energy spectra agree well with the data.
For the $B(M1)/B(E2)$ ratios, a strong staggering behavior is exhibited for bands D3 and D3-C, and the data are somewhat underestimated for band D4.

For bands D5 and D5-C, in consistent with the data, the decreasing trend for the energy differences between the partners with spin is reproduced.
For the $B(M1)/B(E2)$ ratios, the calculated results agree well with the data which have been measured for both yrast and yrare bands.

\begin{figure*}[!]
  \centering
  \includegraphics[width=0.7\textwidth]{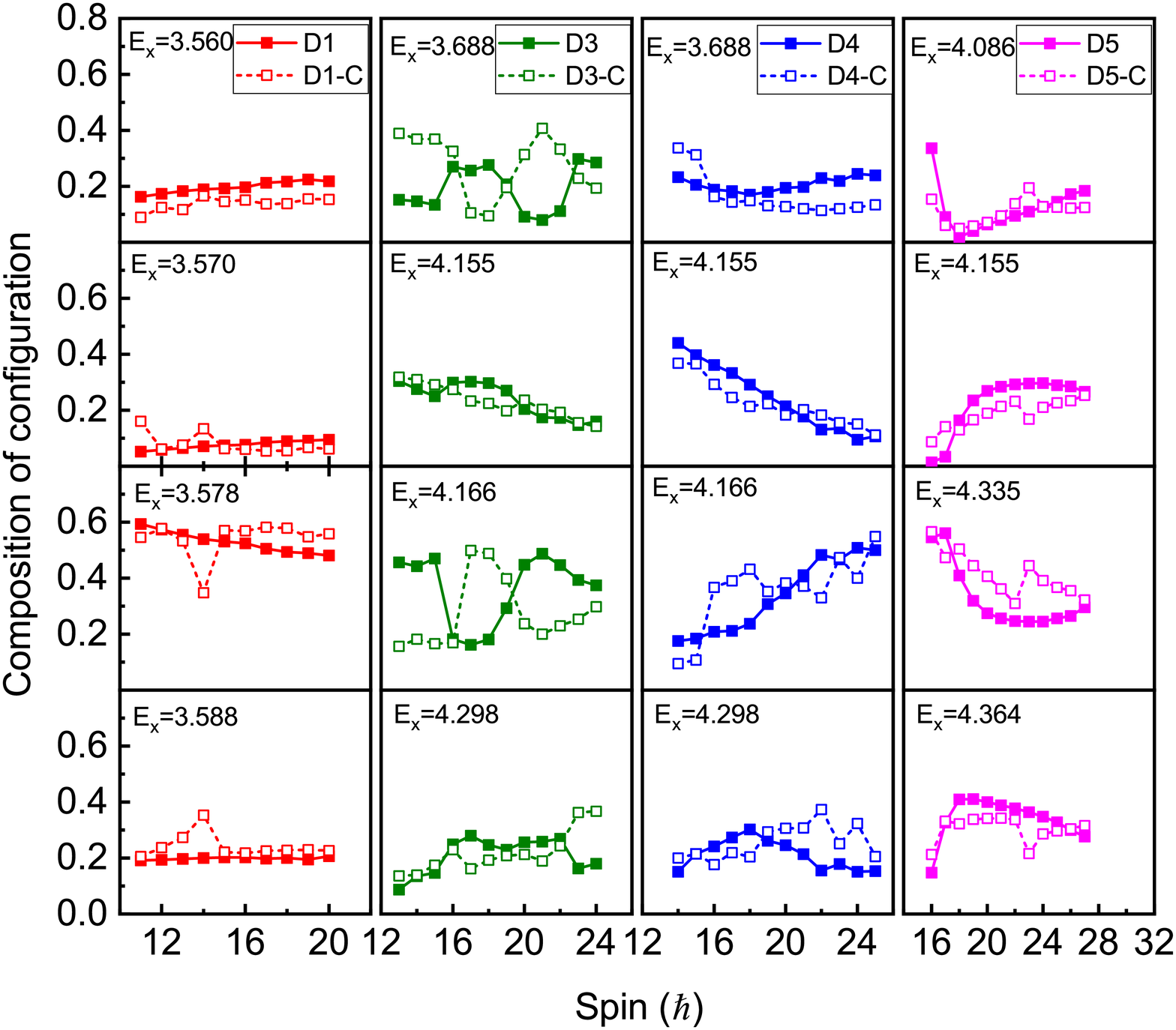}
  \caption{(Color online) The compositions of configurations for bands D1 (1st column), D3 (2nd column), D4 (3rd column), and D5 (last column) as well as for their partners. Each configuration is labeled by the corresponding qp excitation energy $E_x$, and its contribution to the band is obtained by summing over the weights $W_\kappa$ for the $n$-qp states $|\Phi_\kappa\rangle$ with the same qp excitation energy.}
  \label{fig2}
\end{figure*}

From the eigenfunctions in Eq.~\eqref{eq7}, the weight $W_\kappa$ of the $n$-qp state $|\Phi_\kappa\rangle$ can be calculated by Eq.~\eqref{eq8}.
Summing over the $W_\kappa$ for the $n$-qp states with the same qp excitation energy $E_x$ in Table~\ref{table1}, the compositions of configurations for bands D1, D3, D4, D5, and their partners are depicted in Fig.~\ref{fig2}.

For bands D1, D4, and D5, the compositions of configurations are respectively similar to their partner bands, which is in agreement with the expectation of the chiral rotation.
However, the compositions for bands D3 and D3-C exhibit differences for the configurations with $E_x = 3.688$ MeV and $E_x=4.166$ MeV.
These two configurations mainly differ in the quasi-neutron occupation of either $d_{3/2}$ or $d_{5/2}$.
The mixing of these two configurations leads to the rapid change in the compositions, which may explain the strong staggering $B(M1)/B(E2)$ ratios for bands D3 and D3-C shown in Fig.~\ref{fig1}.
The compositions of configurations for bands D2 and D2-C are trivial, since all 32 $6$-qp states have the same qp excitation energy.
Therefore, they are not shown here.

%--------------------------------------------------------------------------
\begin{figure*}[!]
  \centering
  \includegraphics[width=\textwidth]{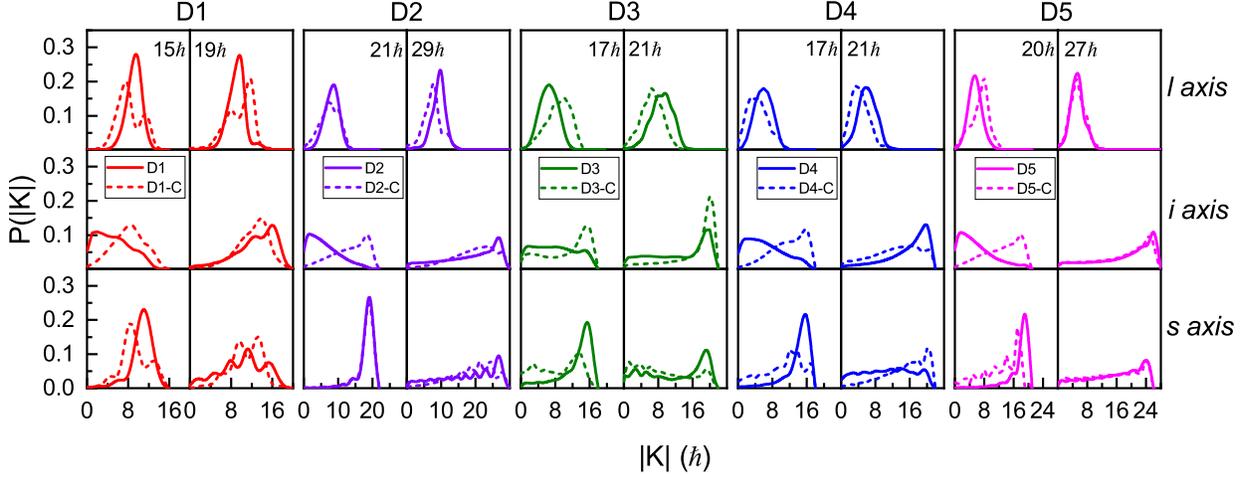}
  \caption{(Color online) The \textit{K plot}, i.e., the $K$ distributions for angular momenta on the three principle axes, for bands D1-D5 and their partners for selected angular momenta.
  The $K$ distributions on the $l$, $i$ and $s$ axes are shown in the first, second and third rows, respectively.}
  \label{fig3}
\end{figure*}

The chiral geometry can be illustrated clearly by the \textit{K plot} and the \textit{azimuthal plot}~\cite{Chen2017Phys.Rev.C96051303,Chen2018PhysicsLettersB785211-216}.
In Fig.~\ref{fig3}, the \textit{K plots}, i.e., the $K$ distributions of the angular momenta along the three principle axes, are given for the five chiral doublets candidates in $^{136}$Nd.

For bands D1 and D1-C, the $K$ distributions at $I = 15\hbar$ and $19\hbar$ are given in the left two columns of Fig.~\ref{fig3}.
The main change of the $K$ distributions occurs on the \textit{i} axis.
At $I=15\hbar$, the $K$ distribution for band D1 along \textit{i} axis peaks at $K=0\hbar$, in contrast to the vanishing $K$ distribution for band D1-C.
This is the typical feature of zero- and one-phonon states, which is interpreted as the chiral vibration with respect to the \textit{l-s} plane~\cite{Qi2009PhysicsLettersB675175-180}.
At $I=19\hbar$, $K$ distributions peak at $K=15\hbar$ for both bands D1 and D1-C.
This suggests that the collective rotation around the $i$-axis develops and the angular momenta deviate from the \textit{l-s} plane.
The similar $K$ distributions for bands D1 and D1-C suggest the occurrence of the static chirality.
Similarly, the evolutions from chiral vibration to static chirality can be found in Fig.~\ref{fig3} for bands D2 and D2-C, D4 and D4-C, and D5 and D5-C.

For bands D3 and D3-C, however, the pattern is slightly different.
As shown in Fig.~\ref{fig3}, at $I = 17\hbar$, chiral vibration with respect to the \textit{l-s} plane disappears due to the strong configuration mixing.
At $I = 21\hbar$, the $K$ distributions exhibit similarity, which might suggest the appearance of static chirality.
The examination of the \textit{azimuthal plot} can further clarify the corresponding chiral geometry.

%---------------------------------------------------------------------------------
\begin{figure*}[!tbp]
  \centering
  \includegraphics[width=0.8\textwidth]{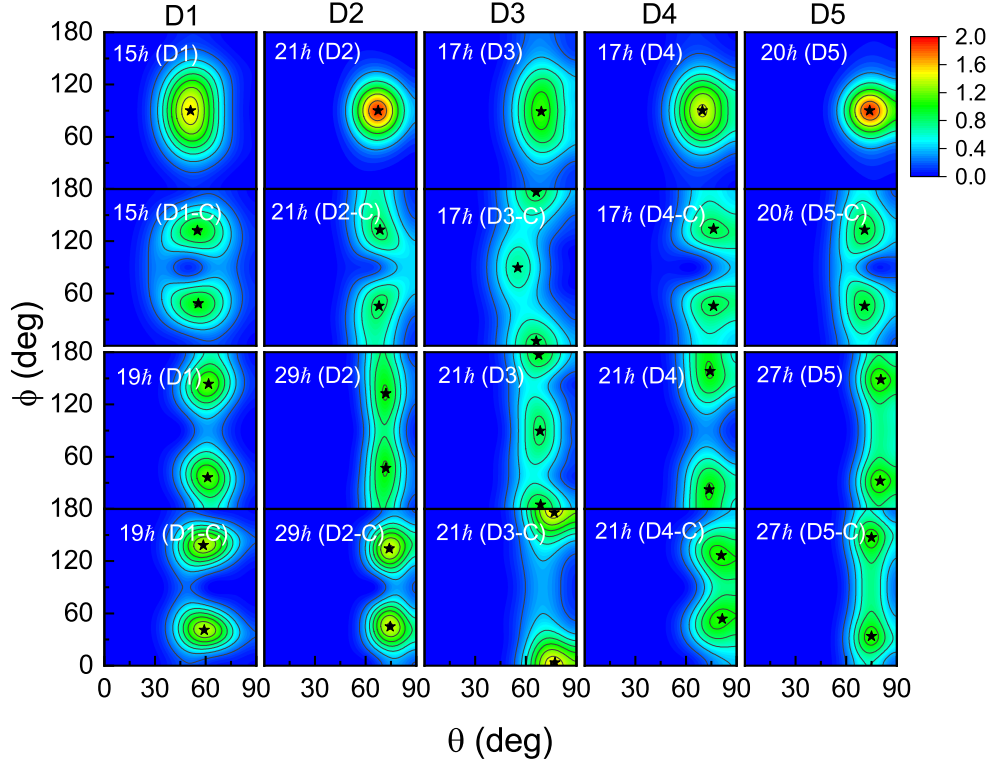}
  \caption{(Color online) The \textit{azimuthal plot}, i.e., probability distribution profiles for the orientation of the angular momentum on the $(\theta,\phi)$ plane for bands D1-D5 and their partners for selected angular momenta.
  The black star represents the position of a local maximum.}
  \label{fig4}
\end{figure*}

In Fig.~\ref{fig4}, the probability distribution profiles for the orientation of the angular momenta on the $(\theta,\phi)$ plane, namely the \textit{azimuthal plots}~\cite{Chen2017Phys.Rev.C96051303,Chen2018PhysicsLettersB785211-216}, for bands D1-D5 and their partners are shown with the same spin as in Fig.~\ref{fig3}.

For bands D1 and D1-C, the \textit{azimuthal plots} at $I = 15\hbar$ and $19\hbar$ are shown in the first column of Fig.~\ref{fig4}.
At $I = 15\hbar$, the \textit{azimuthal plot} for band D1 has a single peak at $(\theta=51^\circ, \phi=90^\circ)$, which indicates the orientation of the angular momentum mainly in the \textit{l-s} plane and corresponds to a planar rotation.

The \textit{azimuthal plot} for band D1-C exhibits two peaks at $(\theta=55^\circ, \phi=48^\circ)$ and $(\theta=55^\circ, \phi = 132^\circ)$.
Therefore, the states at $I = 15\hbar$ for bands D1 and D1-C can be recognized as zero- and one-phonon states.
Same interpretation of chiral vibration as in the \textit{K plot} is obtained.
At $I = 19\hbar$, two peaks at ($\theta = 61^\circ, \phi = 37^\circ$) and ($\theta=61^\circ, \phi=143^\circ$) appear in the \textit{azimuthal plot} for band D1 which correspond to aplanar orientations.
Similar \textit{azimuthal plots} for bands D1 and D1-C suggest the realization of static chirality, which is consistent with the interpretation as in the \textit{K plot}.
Similar interpretation from chiral vibration to static chirality can be found from the \textit{azimuthal plots} in Fig.~\ref{fig4} for bands D2 and D2-C, D4 and D4-C, and D5 and D5-C.

The \textit{azimuthal plots} for bands D3 and D3-C at $I = 17\hbar$ and $21\hbar$ are shown in the third column of Fig.~\ref{fig4}.
At $I = 17\hbar$, the \textit{azimuthal plot} for band D3 has a single peak
at ($\theta = 69^\circ, \phi = 90^\circ$), which means the angular momentum orientates mainly in \textit{l-s} plane and corresponds to a planar rotation.
The \textit{azimuthal plot} for band D3-C exhibits three peaks at $(\theta = 66^\circ, \phi = 0^\circ)$, $(\theta=55^\circ, \phi = 90^\circ)$, and $(\theta=66^\circ, \phi=180^\circ)$.
The first and third peaks represent the same orientation for the angular momentum, and correspond to a planar rotation in the \textit{l-i} plane.
The second peak corresponds to a planar rotation in the \textit{l-s} plane.
At $I = 21\hbar$, three peaks at $(\theta = 68^\circ, \phi = 0^\circ)$, $(\theta = 68^\circ, \phi = 90^\circ)$, and $(\theta = 68^\circ, \phi = 180^\circ)$ for band D3 and two peaks at $(\theta = 77^\circ, \phi = 0^\circ)$ and $(\theta = 77^\circ, \phi = 180^\circ)$ for band D3-C are found in the \textit{azimuthal plot}.
From the features in the \textit{azimuthal plot}, it is difficult to interpret bands D3 and D3-C as chiral partners.

%-----------------------------------------------------------------------------
In summary, a triaxial projected shell model including configurations with more than four quasiparticles in the configuration space is developed, and applied to investigate the recently reported five chiral doublets candidates in a single even-even nucleus $^{136}$Nd.
The energy spectra and $B(M1)/B(E2)$ ratios are reproduced satisfactorily.
The configuration mixing along the rotational bands is studied by analyzing the intrinsic composition of the eigenfunctions.
The chiral geometry of these bands is examined by the \textit{K plot} and the \textit{azimuthal plot}, and the evolutions from chiral vibration to static chirality with spin for bands D1 and D1-C, D2 and D2-C, D4 and D4-C, and D5 and D5-C are clearly demonstrated.
From the features in the \textit{azimuthal plot}, it is difficult to interpret bands D3 and D3-C as chiral partners.
In Ref.~\cite{Chen2018PhysicsLettersB782744-749}, the calculations of particle rotor model have been performed for configurations with three single-$j$ shells to describe D3 and four single-$j$ shells to describe D4.
Except for band D4 with $I\geq 19\hbar$, the calculations of particle rotor model do not agree well with the data.
Future microscopic studies for bands D3 and D3-C, e.g., three-dimensional tilted axis cranking covariant density functional theory~\cite{Zhao2017PhysicsLettersB7731-5}, would be interesting.

%--------------------------------------------
\begin{acknowledgments}
  This work was partly supported by the National Key R\&D Program of China (Contract No. 2018YFA0404400 and No. 2017YFE0116700), the National Natural Science Foundation of China (Grants No. 11621131001 and No. 11875075).
\end{acknowledgments}

%-------------------------------------------------------------------------------------------
%
%-------------------------------------------------------------------------
\end{document}